\documentclass[aps,prl,twocolumn,superscriptaddress]{revtex4}
\usepackage{amssymb,amsmath}
\usepackage{graphicx}

\begin{document}
\title{Pseudogap-less high T$_{\text{c}}$ superconductivity in BaCo$_{\text{x}}$Fe$_{\text{2-x}}$As$_{\text{2}}$}

\author{F. Massee}
\affiliation{Van der Waals-Zeeman Institute, University of Amsterdam, 1018XE Amsterdam, The Netherlands}
\author{Y.K. Huang}
\affiliation{Van der Waals-Zeeman Institute, University of Amsterdam, 1018XE Amsterdam, The Netherlands}
\author{J. Kaas}
\affiliation{Van der Waals-Zeeman Institute, University of Amsterdam, 1018XE Amsterdam, The Netherlands}
\author{E. van Heumen}
\affiliation{Van der Waals-Zeeman Institute, University of Amsterdam, 1018XE Amsterdam, The Netherlands}
\author{S. de Jong}
\affiliation{Van der Waals-Zeeman Institute, University of Amsterdam, 1018XE Amsterdam, The Netherlands}
\affiliation{SLAC National Accelerator Center, Stanford University, Menlo Park, California 94025, United States}
\author{R. Huisman}
\affiliation{Van der Waals-Zeeman Institute, University of Amsterdam, 1018XE Amsterdam, The Netherlands}
\author{H. Luigjes}
\affiliation{Van der Waals-Zeeman Institute, University of Amsterdam, 1018XE Amsterdam, The Netherlands}
\author{J.B. Goedkoop}
\affiliation{Van der Waals-Zeeman Institute, University of Amsterdam, 1018XE Amsterdam, The Netherlands}
\author{M.S. Golden}
\affiliation{Van der Waals-Zeeman Institute, University of Amsterdam, 1018XE Amsterdam, The Netherlands}

\begin{abstract}
The pseudogap state is one of the peculiarities of the cuprate high temperature superconductors. Here we investigate its presence in BaCo$_{\text{x}}$Fe$_{\text{2-x}}$As$_{\text{2}}$, a member of the pnictide family, with temperature dependent scanning tunneling spectroscopy. We observe that for under, optimally and overdoped systems the gap in the tunneling spectra always closes at the bulk T$_{\text{c}}$, ruling out the presence of a pseudogap state. For the underdoped case we observe superconducting gaps over large fields of view, setting a lower limit of tens of nanometers on the length scale of possible phase separated regions.
\end{abstract}
\maketitle

With the discovery of the pnictide family of high temperature superconductors \cite{kamihara}, a new window to the physics behind high temperature superconductivity (HTSC) has been opened. At first glance, the phase diagrams of the cuprates and the pnictides are strikingly similar. In both systems, electron or hole doping suppresses the magnetic ground state of the parent compound and produces a superconducting dome, see for instance \cite{Kivelson_NMat}. However, whereas the cuprates are Mott insulators at low dopings and upon doping first completely lose their long range magnetic order before superconductivity emerges, the pnictide parent compounds are metals and the doped systems can directly cross from a long range ordered magnetic to a superconducting phase upon cooling. The nature of this transition is still under debate, as some studies find that the two regions can coexist \cite{Julien, Drew_NMat_8, kim_JPCM_21} while others indicate a phase separation scenario \cite{Park_PRL_102, takeshita_NJP_11}.

At low doping concentrations, the cuprates furthermore display a second energy scale, characterized in both the charge and spin sectors by a depression of spectral weight at the Fermi level which is observed both in the superconducting and normal state: the so called pseudogap \cite{pseudogap, timusk}. Since the role of the pseudogap phase in the cuprates is still under heavy debate, the establishment of the presence or absence of such a phase in the iron pnictides is of particular importance. There are several indications that at least some of the pnictides have a pseudogapped region at the underdoped side of the phase diagram. For instance, anomalous resistivity characteristics in LaFeAsO$_{\text{1-x}}$F$_{\text{x}}$ and SmFeAsO$_{\text{1-x}}$F$_{\text{x}}$ have been interpreted in a pseudogap-like scenario \cite{buchner_epl}. Nuclear magnetic resonance (NMR) investigations \cite{nakai, grafe} and angle resolved photoemission (ARPES) measurements \cite{sato} have also found signs of a pseudogap in the 1111 systems. In Ba$_{\text{1-x}}$K$_{\text{x}}$Fe$_{\text{2}}$As$_{\text{2}}$ ARPES \cite{ding_arxiv}, optical conductivity \cite{1007.3617} and time resolved, pump-probe spectroscopy \cite{torchinsky_PRL_2010} measurements showed pseudogap-like behavior, and claims of possible pseudogap behavior have also been made in the FeSe$_{\text{1-x}}$Te$_{\text{x}}$ system \cite{arxiv:1001.3273}. In the case of BaCo$_{\text{x}}$Fe$_{\text{2-x}}$As$_{\text{2}}$, ARPES measurements have reported a slight depression in signal above the superconducting transition temperature \cite{ARPES_co}, possibly caused by a weak pseudogap. 

A further experimental observation that could be linked to the presence of a pseudogap phase is the large spread in the peak-to-peak separation (2$\Delta_{\text{p-p}}$) seen in tunneling spectra of the pnictides \cite{Yin_PRL, Massee_PRB_79}. Such a large 2$\Delta_{\text{p-p}}$ variation is difficult to reconcile with the sharpness of the superconducting transition measured using resistive or magnetic means. In underdoped cuprates, the analogous tunneling spectra are dominated by pseudogap features, and the true superconducting quasiparticle spectrum has been resolved only very recently \cite{Boyer_Nature}. The question is therefore whether the tunneling spectra of the pnictide HTSC also show a mix of pseudogap and superconducting gap features.

\begin{figure*}[t]
\includegraphics[width=2\columnwidth]{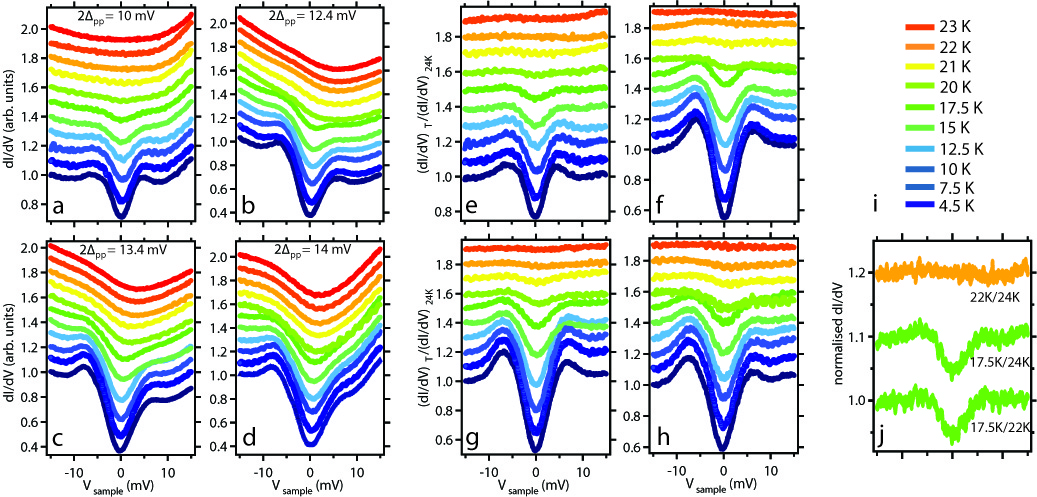}
\caption{\label{FIG:opt}Optimally doped BaCo$_{\text{x}}$Fe$_{\text{2-x}}$As$_{\text{2}}$ (x=0.14, T$_{\text{c}}$ = 22 $\pm$ 0.5 K). Panels (a)-(d): differential conductance (LDOS) traces for four different superconducting gap sizes, as a function of temperature. For each panel, the T-dependent data were recorded from exactly the same real space location. The spectra are offset vertically for clarity. V = 45 mV, I = 40 pA. Panels (e)-(h): the same spectra, but now normalized to the T=24K spectrum for each location.  (i) color legend for (a)-(h). (j) An example of a comparison between different normalization temperatures, illustrating that as long at the normalization spectrum is recorded above T$_c$, the results remain unaltered.}
\end{figure*}

In this Letter we report temperature and doping dependent scanning tunneling spectroscopy measurements on underdoped, optimally and overdoped BaCo$_{\text{x}}$Fe$_{\text{2-x}}$As$_{\text{2}}$ crystals (x=0.08, 0.14 and 0.21). The main goal is to determine whether all the gaps seen in tunneling experiments are indeed superconducting gaps that close at T$_c$, or whether there are also pseudogapped tunneling signatures for which the gap does not close for T$>$T$_c$. There is some debate in the literature on the nature of the cleavage surface of the 122-pnictide family \cite{Boyer, Yin_PRL, Yazdani, Massee_PRB_79, Plummer, Masseecleavage, Chinesegroup}. Our view is that, in order to avoid creation of a polar surface, the crystal cleaves in the Ba layer, exposing half a Ba layer on either side of the cleave \cite{Masseecleavage, Chinesegroup, surfacecalc, heumen-arxiv}. Although a comparison of VUV and hard x-ray photoemission data suggests that the cleavage surface does not have a dominating  influence on the global near-surface electronic environment \cite{SdJ_PRB_HIKE}, LDA calculations report electronic properties differing from the bulk for $\sqrt{2}\times\sqrt{2}$ and $2\times1$ reconstructed Ba surfaces \cite{Chinesegroup, heumen-arxiv}. However, as we observe identical peak-to-peak gaps in tunneling from both $\sqrt{2}\times\sqrt{2}$ as well as $2\times1$ or other low-corrugation surface terminations, we present data here from spectroscopic surveys recorded on the modal termination topography. We do note that areas with relatively large z-corrugation ($>$2 \AA) show anomalous spectroscopic signatures, which we attribute to local Ba surplus or deficiency. Such regions of the cleavage surfaces were not used for the spectroscopic surveys reported here. The STM/STS microscope, which enables us to track the surface with atomic precision as a function of temperature, has been described elsewhere \cite{Massee_PRB_79}. 

\begin{figure}[h]
\includegraphics[width=8cm]{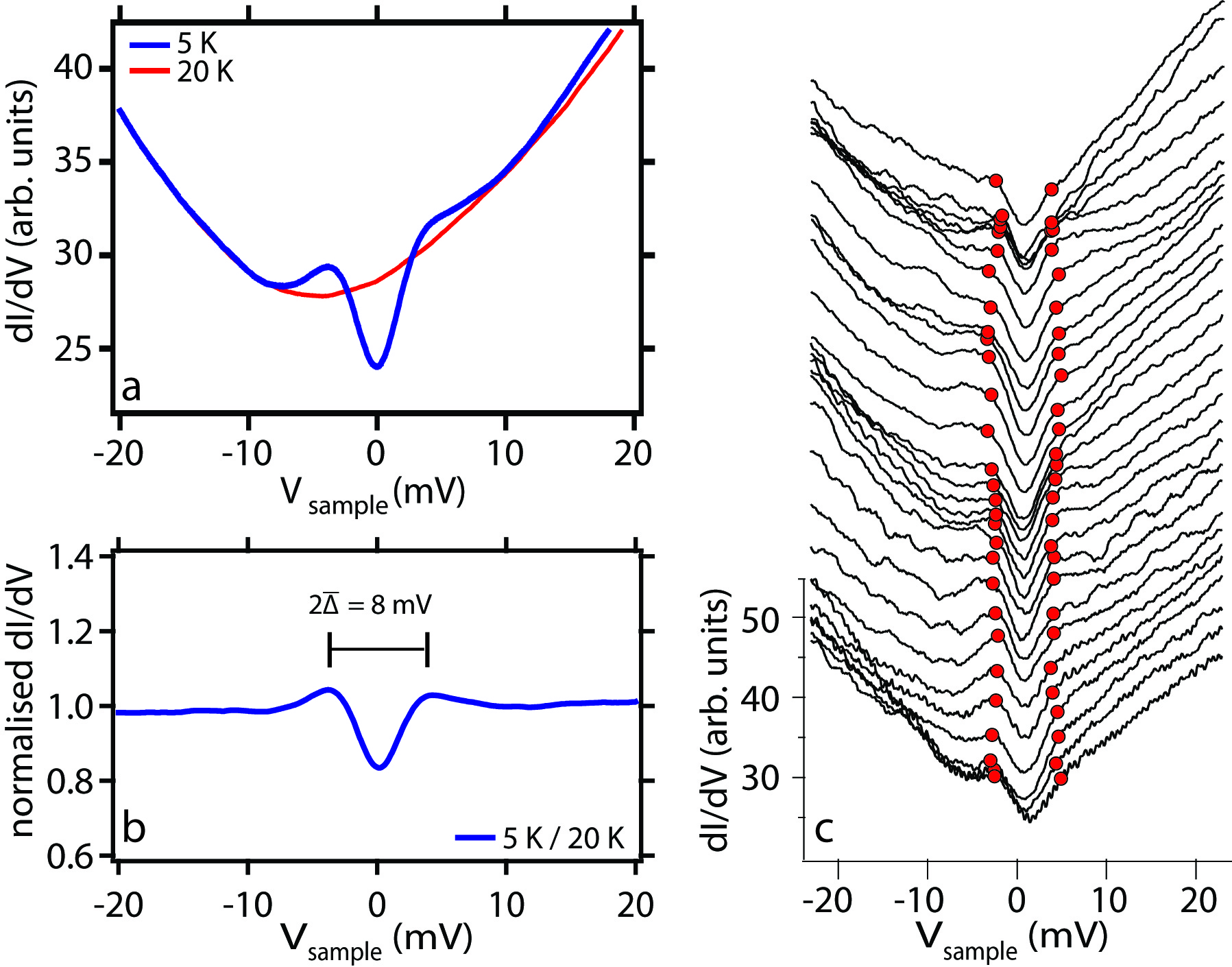}
\caption{\label{FIG:underdoped} Underdoped BaCo$_{\text{x}}$Fe$_{\text{2-x}}$As$_{\text{2}}$ (x=0.08, T$_{\text{c}}$ = 14 $\pm$ 1 K): (a) Average of $\sim$4000 LDOS spectra taken both at 5 K and 20 K on a square grid. Both the 5K and 20K data were recorded on the same 100x100 \AA$^{2}$ field of view. (b) Spectrum obtained by normalizing the 5 K trace shown in (a) to the 20 K data. (c) A collection of single pixel spectra taken along a linesman through the conduction map whose average is shown in (a), illustrating the point to point variation in the spectra. The gap edges are indicated with circles.}
\end{figure}

\begin{figure*}[ht]
\includegraphics[width=2\columnwidth]{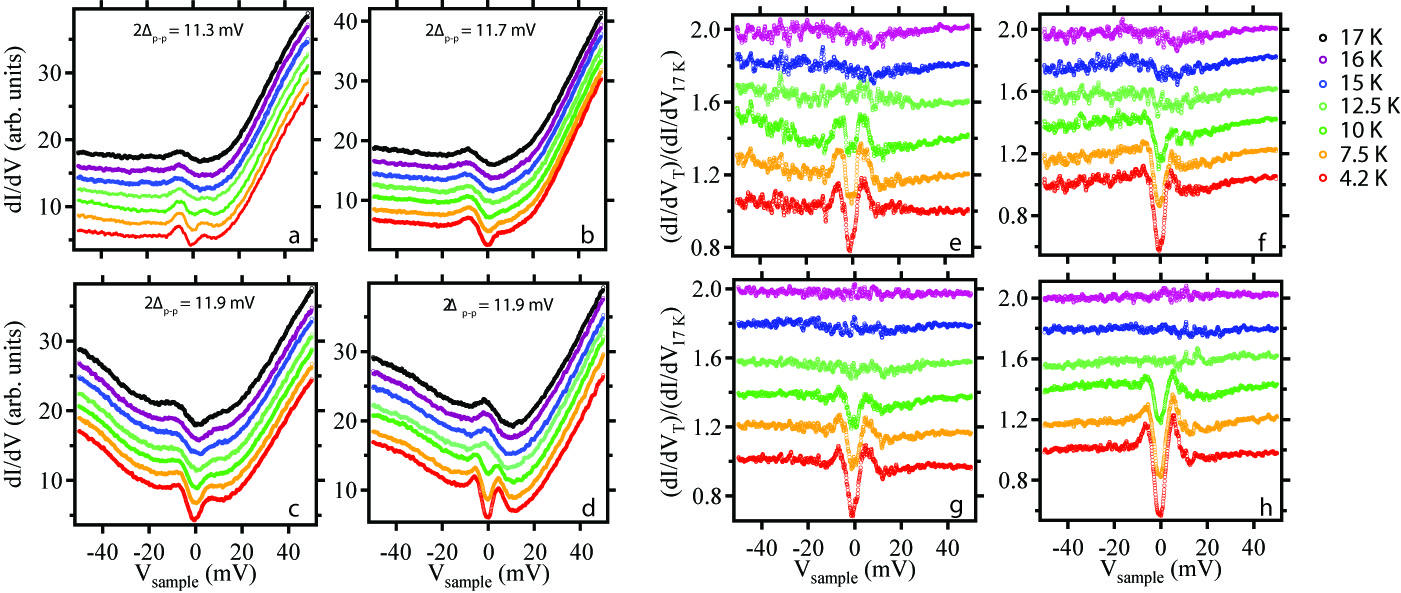}
\caption{\label{FIG:fig3}Overdoped BaCo$_{\text{x}}$Fe$_{\text{2-x}}$As$_{\text{2}}$ (x=0.21, T$_{\text{c}}$ = 13 $\pm$ 1 K): (a)-(d) Temperature dependent LDOS spectra recorded at four different real space locations. The data for T$>$T$_c$ for the different locations fall perfectly on top of each other, contrary to the data for T$<$T$_c$. Panels (e)-(h) show the spectra (a)-(d) normalized to the respective spectra taken at 17 K, i.e. in the normal state.}
\end{figure*}

We begin by presenting the data from optimally doped BaCo$_{\text{x}}$Fe$_{\text{2-x}}$As$_{\text{2}}$ (x = 0.14) samples which have T$_{\text{c}}$ = 22 $\pm$ 0.5 K. 
Figure 1(a)-(d) shows the evolution of the gapped LDOS for four different locations on the surface with temperatures ranging from 4.5 K to 24 K. As can be seen in the figure, the clear coherence peaks recorded for T=4.5 K are smoothly suppressed and are undetectable at temperatures above T$_{\text{c}}$. The U-shaped background of the spectra makes it difficult to follow the temperature dependence in a quantitative manner. Consequently, we adopt the common procedure of dividing the differential conductance at a temperature T, by that taken in the normal state, i.e. (dI/dV)$_{\text{T}}$/(dI/dV)$_{\text{T}_{\text{normal}}}$. The result of this analysis is plotted in Fig. 1(e)-(h). Here, unlike in the cuprates, the gap spectrum obtained after normalization shows the same 2$\Delta_{\text{p-p}}$ as is apparent in the raw data in the superconducting state for all four magnitudes of the peak-to-peak gap sampled. The exact temperature at which the gap closes, or fills-in, is difficult to estimate, but for all gap sizes it is within 1 K of the bulk T$_{\text{c}}$ determined from resistivity measurements from the same crystals.

To ensure that the choice of normalization temperature used in the procedure described above does not influence the conclusions, normalization of a spectrum taken at 17.5 K to both the 22 K data, i.e. very close to T$_{\text{c}}$, and those recorded at 24 K are compared in Fig. 1(j). The result of both normalizations is so similar that we can safely say that the results of the normalization are robust, as long a T$>$T$_{\text{c}}$ is chosen for the division. Summarizing for the optimally doped system, it is thus safe to say that the gaps observed in our STS data are indeed superconducting gaps. 

We now turn our attention to lower doping concentrations, where, in analogy with the cuprates, it could be expected that pseudogap behavior would show up more strongly, mediated -  possibly - by enhanced magnetic correlations. Figure 2(a) shows the average of $\sim$4000 spectra taken at 5 K and 20 K on the same 100x100 \AA$^{2}$ field of view on a BaCo$_{\text{x}}$Fe$_{\text{2-x}}$As$_{\text{2}}$ crystal with x=0.08 (T$_{\text{c}}$ = 14 $\pm$ 1 K, T$_{\text{N}}$ = 70 $\pm$ 3 K). The low temperature spectrum, normalized to the normal state spectrum, is shown in Fig. 2(b).

One peculiarity valid for all our measurements of underdoped crystals is that the gap signatures are less pronounced than in the optimally doped materials. Nevertheless, the single pixel spectra, of which several are shown in Fig. 2c, clearly show there are coherence-peak like gaps all over the field of view. As for the optimally doped case, the peak-to-peak separation does not change on normalization, and the gaps all close at the bulk T$_{\text{c}}$. Again, these observations strongly support the notion that the tunneling gaps observed at low temperature are indeed superconducting gaps and not gaps showing the same phenomenology as the pseudogaps in cuprate superconductors \cite{Boyer_Nature}. 

Since at 20 K the underdoped system is still well within the orthorhombic, magnetically ordered state, a certain suppression of the LDOS around E$_F$ could be expected due to the gapping of significant parts of the Fermi surface observed in quantum oscillation and ARPES experiments \cite{sebastian_jpcm_2008, analytis_prb_2009, dejong_epl_2010}. For a mean field gap linked to the structural/magnetic ordering temperature, such features could be $\sim$10 mV away from zero bias. Our STS data show no signs of such an LDOS suppression, which may be due either to tunneling matrix element effects or to the difference in LDOS enhancement at the gap edges between partial gapping in $k$-space due to back-folding and hybridization of bands in the magnetically ordered, orthorhombic state and the coherence peaks of the superconducting state. Additional measurements, for instance to higher energies, should be performed to address this point in a conclusive manner. In any case, the presence of superconducting gaps over the entire field of view sets a lower limit of tens of nanometers on the length scale of possible phase separation between superconducting and non-superconducting regions of the sample surface.

To complete our survey of the phase diagram of BaCo$_{\text{x}}$Fe$_{\text{2-x}}$As$_{\text{2}}$, we turn our attention to the overdoped compound (x=0.21, T$_{\text{c}}$ = 13 $\pm$ 1 K). In analogy with Fig. 1, Fig. 3(a)-(d) shows the detailed temperature dependence of four locations with different 2$\Delta_{\text{p-p}}$. Figure 3(e)-(h) show the traces after  normalization to a normal state spectrum recorded from the same real-space location. As was the case for the under and optimally doped samples, the gaps seen in the normalized  T$<$T$_c$ spectra have identical magnitude to those in the raw data, and all vanish above the bulk T$_{\text{c}}$.

To summarize, for underdoped, optimally doped, and overdoped BaCo$_{\text{x}}$Fe$_{\text{2-x}}$As$_{\text{2}}$ (x=0.08, 0.14 and 0.21), the peak-to-peak gap in the local tunneling density of states, 2$\Delta_{\text{p-p}}$, has been tracked as a function of location on the surface and as a function of temperature. The gaps for all doping concentrations studied and all gap sizes observed vanish at the bulk T$_{\text{c}}$, excluding a cuprate-like pseudogap scenario in these pnictide superconductors. Since superconducting gaps have been seen across the entire field of view of low-corrugation surfaces over areas of hundreds of square \AA ngstrom, nano-scale phase separation of non-superconducting (magnetic) and superconducting patches seems to be unlikely, also in the underdoped material.

The experimental indications of the presence of a pseudogap in several compounds \cite{buchner_epl, nakai, grafe, sato, ding_arxiv,1007.3617, torchinsky_PRL_2010,arxiv:1001.3273} closely related to the one studied here makes the family of pnictide superconductors a theoretical challenge to understand, as the action of specific dopant atoms clearly can lead to completely differing behavior. 
For the Ba122 systems, one could draw a parallel with the cuprates, in which the electron-doped  systems seem less prone to pseudogap physics than the p-type \cite{0906.2931}, although the different behavior for the 1111 pnictide systems again underlines that subtle differences in doping route and crystal chemistry can lead to large differences in the observed physical phenomenology.
Consequently, our new data presented here set a first step with this clear result in the Co-doped Ba122 system. Mapping out the presence or absence of a pseudogap in other related pnictide compounds -  both electron and hole doped - will be an important next step in gaining a better understanding of the role of the pseudogap for high temperature superconductivity in the iron and copper-based materials.

\section{Acknowledgements}
We would like to acknowledge J. S. Agema for expert technical support. This work is part of the research program of FOM (09PR2657), which is financially supported by the NWO.

\end{document}